\documentclass[aps,prb,showpacs,tightenlines,twocolumn]{revtex4-1}

\usepackage{amsmath,amssymb,amsfonts,bm}
\usepackage{graphicx}
\usepackage{dcolumn}
\usepackage[colorlinks=true,linkcolor=blue]{hyperref}%

\begin{document}

%-----------------------------------------------------------------
% documentation    title,  authors,   abstract,   pacs
%-----------------------------------------------------------------

\title{Spin-spin interaction in the bulk of topological insulators}

\author{Jinhua Sun$^{1,2}$,  Liang Chen$^{1}$ and Hai-Qing Lin$^{1}$}

\affiliation{
$^1$Beijing Computational Science Research Center, Beijing, 100084,
People's Republic of China\\
$^2$School of Physics, Peking University, Beijing, 100871,
People's Republic of China
}

\date{\today}

\begin{abstract}
  We apply mean-field theory and Hirsch-Fye quantum Monte Carlo method to study the
  spin-spin interaction in the bulk of three-dimensional topological insulators. We find that
  the spin-spin interaction has three different components: the longitudinal,
  the transverse and the transverse Dzyaloshinskii-Moriya-like
  terms. When the Fermi energy is located in the bulk gap of topological insulators, the
  spin-spin interaction decays exponentially due to Bloembergen-Rowland interaction.
  The longitudinal correlation is antiferromagnetic and the transverse correlations are
  ferromagnetic. When the chemical potential is in the conduction or valence band,
  the spin-spin interaction follows power law decay, and isotropic ferromagnetic
  interaction dominates in short separation limit.
\end{abstract}

\pacs{75.30.Hx, 75.10.-b, 75.40.Mg, 76.50.+g}

\maketitle

%-----------------------------------------------------------------
% The body of the paper
%-----------------------------------------------------------------

\section{Introduction}

Time-reversal invariant topological insulators{\cite{PhysRevLett.98.106803,Nat.Phys.5.398,
Nat.Phys.5.438,Nature.464.194,RevModPhys.82.3045,RevModPhys.83.1057}} (TIs) have attracted
much attention and have been extensively investigated in the past few years. Three-dimensional
(3D) TIs are fully gapped in the bulk with strong spin-orbit coupling, but have protected gapless
surface states. Semiconducting thermoelectric Bi$_2$Se$_3$, Bi$_2$Te$_3$ and Sb$_2$Te$_3$
are most studied promising materials of TIs with large bulk band gaps $\sim$ 0.3eV and gapless
Dirac fermions (quasi-particles) on the surface
.{\cite{Nat.Phys.5.438,RevModPhys.82.3045,RevModPhys.83.1057}}
There are many interesting
phenomena related to spin-orbit locked TI surface states{\cite{PhysRevB.78.195424}},
including the topological magnetoelectric effect{\cite{PhysRevB.78.195424}}, the half integer
quantum Hall effect {\cite{PhysRevLett.106.166802}}, the image magnetic monopole
{\cite{Qi27022009,PhysRevB.81.245125}} induced by electric charge near the TI surface state,
the topologically quantized magneto-optical Kerr and Faraday rotation
{\cite{PhysRevLett.105.057401,PhysRevB.84.205327}} in units of the fine structure constant
{\cite{PhysRevLett.105.166803,PhysRevB.83.205109}} and the repulsive Casimir effect between
TIs with opposite topological magnetoelectric polarizabilities
{\cite{PhysRevLett.106.020403,PhysRevB.84.045119,PhysRevB.84.075149}}, {\it etc}..

To study the various phenomena mentioned above and control the transport properties of TIs,
one needs to break the time-reversal symmetry and generate an energy gap for TI surface Dirac
fermions. Surface- and bulk-doping with magnetic impurities are feasible methods.
{\cite{Chen06082010,Wray07322011,PhysRevLett.106.206805,PhysRevLett.108.117601,PhysRevB.81.195203,
PhysRevLett.108.256810}} There are numerous works demonstrating that magnetic doping can induce
a surface band gap.{\cite{Chen06082010,Wray07322011,PhysRevLett.106.206805}} However, some
other experiments could not observe the gap opened by magnetic doping
{\cite{PhysRevLett.108.117601,PhysRevB.81.195203,
PhysRevLett.108.256810}}, such that the Ruderman-Kittel-Kasuya-Yosida (RKKY){\cite{Kasuya01071956,PhysRev.96.99,PhysRev.106.893}} interactions between
magnetic impurities have attracted much attention. For the surface-doping
{\cite{PhysRevLett.102.156603}}, careful theoretical investigations
{\cite{PhysRevLett.106.097201,PhysRevLett.106.136802,PhysRevB.81.233405}} show that the RKKY
interactions mediated by surface Dirac fermions are much complicated than expected. Other than the
normal Heisenberg-like interactions, there may exist Ising-like and Dzyaloshinskii-Moriya
(DM)-like interactions between magnetic impurities on the surface of TIs. For the bulk-doping
{\cite{PhysRevB.87.195122}}, mean-field analysis {\cite{PhysRevB.85.195119}} shows that there
may exist ferromagnetic (FM) or antiferromagnetic (AFM) correlation between magnetic impurities.
Density functional theory calculations{\cite{PhysRevB.84.245418,PhysRevLett.108.206801,
PhysRevLett.109.266405}} demonstrate complicated anisotropic spin texture in magnetically doped
TIs. However, the importance of spin-orbit coupling in RKKY interactions has not been
clarified. A recent experiment{\cite{PhysRevLett.110.136601}} sheds light on complicated
RKKY interactions in the bulk of TIs. It is reported that magnetic impurities in the bulk
of Fe$_x$Bi$_2$Se$_3$ behave like ferromagnetic-cluster glass for $x\sim0.025$, and valence-bond
glass for the region from $x\sim0.03$ up to $x\sim0.1$. A detailed analysis on the spin-spin
interaction in the bulk of TIs may provide clue to these results.

In this paper, we apply mean-field theory and Hirsch-Fye quantum Monte Carlo (HFQMC)
\cite{PhysRevLett.56.2521} method to study the carrier-mediated spin-spin interaction between two magnetic
impurities in the bulk of TIs. To get a heuristic physical picture about the correlation
between magnetic impurities, we use mean field theory to study the problem before applying
the  quantum Monte Carlo method. We take two steps. Firstly, we use the self-consistent
Hartree-Fork approximation to estimate the local moment of a single impurity. The
self-consistent Hartree-Fork approximation gives analytical criterion{\cite{PhysRev.124.41}}
about the formation of local moment in the background of electron gas. It was also applied
to study the local moment problem in other systems ({\it i.e.} semiconductor{\cite{PhysRevB.13.2553}}
and graphene{\cite{PhysRevLett.101.026805}}). Once we get a nonzero magnetic moment, we
use functional integral{\cite{Negelebook}} approach to study the interaction between magnetic
impurities, the approach is a little different from the original RKKY perturbation theory
 because we have to deal with spin-orbit coupled systems.
When quantum fluctuation is significant, the mean-field results are suspectable, so
furthermore, we use the quantum Monte Carlo method to study the problem. The HFQMC technique
is a numerically exact method, it is widely used to study the magnetic properties of impurities
in metals, {\it i.e.} the local moment of impurity{\cite{PhysRevLett.56.2521,PhysRevB.38.433}},
the Kondo effect{\cite{PhysRevB.35.4901}} and the interaction between localized moments
{\cite{PhysRevB.35.4901,PhysRevB.35.4943}}, {\it etc.}. Recently, the HFQMC technique is
applied to study the magnetic properties of Anderson impurities in dilute magnetic
semiconductors{\cite{PhysRevB.76.045220}} and graphene{\cite{PhysRevB.84.075414}}.

The paper is organized as follows. We present the model and mean-field results in Sec.
\ref{section2}. In Subsection \ref{section2.1}, we introduce the model Hamiltonian describing
the magnetic impurities in TIs. In subsection \ref{section2.2}, we present the mean-field
approach and analyze the spin-spin interaction for different chemical potential regions
and impurity energy levels. In Sec. \ref{section3}, we show the results obtained by the HFQMC simulations.
We investigate the local moment in subsection \ref{section3.1}, and the spin-spin interaction in subsection \ref{section3.2}.
Finally, discussions and conclusions are given in
Sec. \ref{section4}.

\section{Model Hamiltonian and mean field results}\label{section2}
\subsection{Model Hamiltonian}\label{section2.1}

\begin{figure}[tb]
  \centering
  \includegraphics[width=\columnwidth]{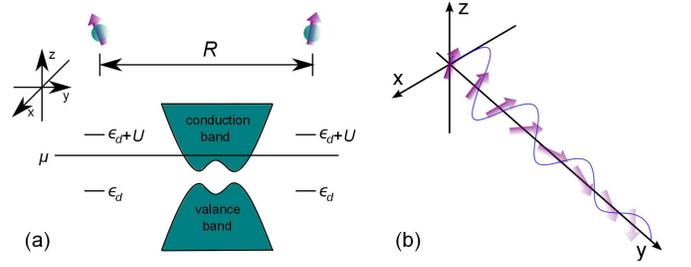}\\
  \caption{(a) Schematic of two magnetic impurities in the bulk of TIs, with RKKY interaction
  mediated by the bulk states of TIs. The two impurities are put on the y-axis, $R$ is the
  distance between impurities. $\epsilon_d$ and $\epsilon_d+U$ are energy levels of impurities,
  the conduction and valence band indicate the band structure of TIs. The chemical potential
  $\mu$ can be tuned into the valence band, conduction band or in the bulk gap by doping.
  (b) Illustration of spin-spin interaction as a function of distance $R$. The blue line means the
  correlation strength and arrows denote the rotation of spin-spin interaction due to the spin-orbit
  coupling in TIs.}\label{fig1}
\end{figure}

We use the four-band model{\cite{PhysRevB.87.195122,PhysRevB.81.115407,1367-2630-12-4-043048,
SHEN01012011}} to describe the bulk states of TIs. The coupling between magnetic
impurities and bulk states can be described by the Anderson model{\cite{Hewsonbook}}. The
total Hamiltonian with TI host material, magnetic impurities and hybridization between TI
bulk states and impurities can be written as
\begin{eqnarray}
\mathcal{H}&=&\mathcal{H}_{\text{b}}+\mathcal{H}_{d}+\mathcal{H}_{\text{hb}},\label{eq1}\\
\mathcal{H}_{\text{b}}&=&\int{\mathbf{d}r}\,\Psi^{\dagger}\left[(m-B{v_F^2}\bm{k}^2)\tau_{z}+
v_F{\bm{k}\cdot\sigma}\tau_{x}-\mu\right]\Psi,\label{eq2}\\
\mathcal{H}_{d}&=&\sum_{j=1,2;s=\uparrow,\downarrow}d_{j,s}^{\dagger}\left(\epsilon_d-\mu\right)d_{j,s}
+U{}d_{j,\uparrow}^{\dagger}d_{j,\uparrow}d_{j,\downarrow}^{\dagger}d_{j,\downarrow},\label{eq3}\\
\mathcal{H}_{\text{hb}}&=&\int{\mathbf{d}}r\,\sum_{j=1,2}{d^{\dagger}_{j}}V_{j}(\bm{r})\Psi(\bm{r})
+h.c., \label{eq4}
\end{eqnarray}
where $\bm{k}=-i\nabla$ is the momentum operator (we take the Planck constant $\hbar=1$),
$\tau_{z,x}$ and $\sigma_{x,y,z}$ are the Pauli matrices for different orbits and different spins,
respectively. $v_F$ is the velocity of the bulk Dirac fermions, and $(m-B{v_F^2}\bm{k}^2)$ is the
effective mass which can be derived from the $k\cdot{p}$ theory{\cite{Nat.Phys.5.438}}. The
sign of $mB$ determines the topology of the bulk states: $mB>0$ ($mB<0$) corresponds to a
topological (normal) insulator. $\mu$ is the chemical potential which can be tuned by doping
{\cite{Hsieh2009}}. The basis vectors of the TI bulk states are chosen as
\begin{equation*}
\Psi=\left(\begin{array}{cccc}
    \psi_{a\uparrow} & \psi_{a\downarrow} & \psi_{b\uparrow} & \psi_{b\downarrow}
\end{array}\right)^{T},
\end{equation*}
where $a$ and $b$ are orbit indices. $d^{\dagger}_{j\uparrow(\downarrow)}$ and
$d_{j\uparrow(\downarrow)}$ are creation and annihilation operators of spin-up (spin-down)
state on the $j$-th impurity site. $(\epsilon_d-\mu)$ is the impurity energy level and $U$ is
the on-site Coulomb interaction. $V_{j}(\bm{r})$ is a $2\times4$ hybridization matrix, which
can be written in the following form in short range coupling limit,
\begin{equation*}
V_{j}(\bm{r})=\left(
    \begin{array}{cccc}
    V_a & 0 & V_b & 0 \\
    0 & V_a & 0 & V_b \\
    \end{array}
\right)\delta(\bm{r}-\bm{r}_j),
\end{equation*}
where $\bm{r}_j$ is the coordinate of the $j$-th impurity. Furthermore, we consider the case
that the impurities are symmetrically coupled with the two orbits, {\it i.e.} $V_a=V_b=V_0$.
Without loss of generality, we assume that the two impurities are put on the $y$-axis and
the distance is $R$. Fig. (\ref{fig1}) gives a schematic of our model.

\subsection{Mean-field results}\label{section2.2}

Before a detailed discussion on the spin-spin interaction, we first make a note
on the units and parameters chosen in this paper. The Planck constant $\hbar$ is set
to be 1, and we choose the parameter $m$ of TIs as energy unit. In the continuous limit, the
summation over momentum is approximated by integration,
$\frac{1}{N_0}\sum_{k}\rightarrow\frac{6\pi^2}{k_T^3}\int\frac{\mathbf{d}^3k}{(2\pi)^3}$,
where $N_0$ is the number of system sites with $k_T$ being the truncation of momentum,
which is determined by the cut-off of the band width
$D=\sqrt{({v_F}{k_T})^2+(m-B{v_F^2}{k_T^2})^2}$. $\Gamma_0=\pi\rho_0{V_0^2}$ represents the
hybridization strength among the impurities and the bulk states, where $\rho_0=N_0/2D$ is
the density of states per spin at the chemical potential. $D=30m$ and $\Gamma_0=0.5$ are
chosen in numerical calculations{\cite{PhysRevB.87.195122}}.

\begin{figure}[tb]
  \centering
  \includegraphics[width=\columnwidth]{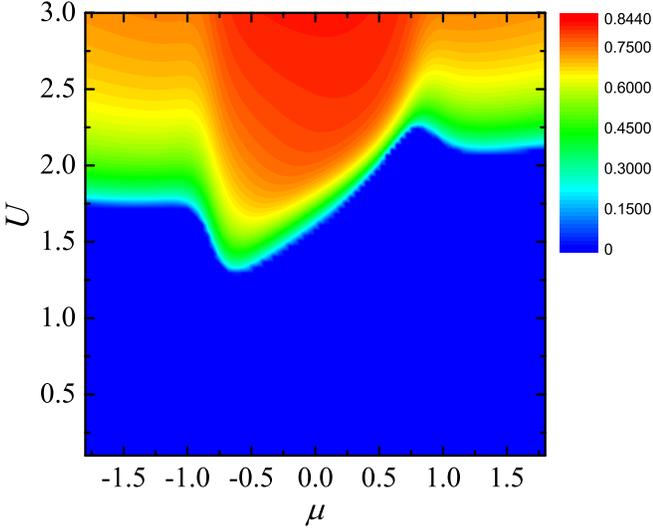}
  \caption{Mean field results of single impurity local moment
  ${m_d}=\langle{n_{\uparrow}}\rangle-\langle{n_{\downarrow}}\rangle$ as a function of
  chemical potential $\mu$ and Hubbard $U$. Impurity energy level $\epsilon_d-\mu=-1.5$
  and temperature $k_BT=1/16$ were chosen in the calculation.}\label{fig2}
\end{figure}

\begin{figure}[tb]
  \centering
  \includegraphics[width=\columnwidth]{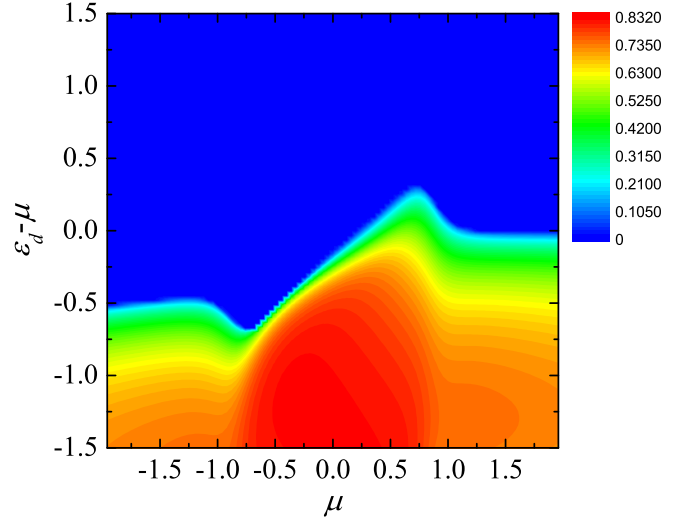}
  \caption{Mean field results of single impurity local moment
  ${m_d}=\langle{n_{\uparrow}}\rangle-\langle{n_{\downarrow}}\rangle$ as a function of
  chemical potential $\mu$ and impurity energy level $\epsilon_d-\mu=-1.5$, Hubbard
  $U=3.0$ and temperature $k_BT=1/16$ were chosen in numerical calculation.}\label{fig3}
\end{figure}

In order to investigate the spin-spin interaction between the two magnetic impurities, a proper
parameter region (the on-site Coulomb repulsion $U$, the impurity energy level $\epsilon_d$ and the chemical potential
$\mu$) are required for the well developed local moment. Under the self-consistent Hartree-Fork approximation, the
imaginary-time Green's function of spin-up and spin-down electrons are
\begin{align}
&g_{\uparrow}(\tau)=\frac{1}{\beta}\sum_{i\omega_n}\frac{e^{-i\omega_{n}\tau}}{i\omega_n-[\epsilon_d-\mu+\Sigma_{1}(i\omega_n)+U\langle{n_{\downarrow}}\rangle]},\label{eq5}\\
&g_{\downarrow}(\tau)=\frac{1}{\beta}\sum_{i\omega_n}\frac{e^{-i\omega_{n}\tau}}{i\omega_n-[\epsilon_d-\mu+\Sigma_{1}(i\omega_n)+U\langle{n_{\uparrow}}\rangle]},\label{eq6}
\end{align}
where $\beta=1/k_BT$, $T$ is the temperature and $k_B$ is the Boltzmann constant.
$\Sigma_{1}(i\omega_n)$ is the self-energy comes from the hybridization between the impurity
and the bulk states, the exact form of the self-energy is
\begin{align}
&\Sigma_{1}(i\omega_n)=-\frac{12i{}D\Gamma_{0}}{k_T^3}\frac{\sqrt{2mB-\lambda-1}-\sqrt{2mB+\lambda-1}}{2\sqrt{2}mB\lambda}, \label{eq7}\\
&\lambda = \sqrt{1+4mB(mB(i\omega_n+\mu)^2-1)},\label{eq8}
\end{align}
$\langle{n_{\downarrow}}\rangle$ and $\langle{n_{\uparrow}}\rangle$ are expectation values
of local spin-up and spin-down states, which can be solved by self-consistent equations
\begin{align}
&\langle{n_{\downarrow}}\rangle=g_{\downarrow}(\tau\rightarrow0^{-}), \label{eq9}\\
&\langle{n_{\uparrow}}\rangle=g_{\uparrow}(\tau\rightarrow0^{-}).\label{eq10}
\end{align}
Fig. (\ref{fig2}) and Fig. (\ref{fig3}) show the numerical results of local moment as a function
of Hubbard $U$, impurity energy level $\epsilon_d-\mu$ and chemical potential $\mu$. One can
find that the local moment $m_d>0.72$ when $\epsilon_d-\mu=-1.5$, $U=3.0$, and $k_BT=1/16$.
Therefore, within the self-consistent mean-field approximation, $U\ge3.0$ is sufficiently
large to generate a non-zero local moment if $\epsilon_d-\mu=-1.5$ and $k_BT\le1/16$.

Now we construct the spin-spin interaction between the local moments, the partition function of
the two impurities in the bulk of TIs can be written as
\begin{align}
&{\cal Z}=\int{{\cal D}d^{\dagger}{\cal D}d}\mathrm{e}^{-\mathcal{S}}, \label{eq11}\\
&\mathcal{S}=-\int_0^{\beta}{\mathrm{d}}\tau{}d^{\dagger}\left[\frac{\partial}{\partial{\tau}}+\left(\epsilon_d-\mu+U{}n_d+\Sigma_1 \right.\right.\nonumber\\
&\left.\left.-U\frac{m_d}{2}{\cal U}\sigma_z{\cal U}^\dagger\right)+(\Sigma_2\lambda_x+\Sigma_3\lambda_y\sigma_y)\right]d, \label{eq12}
\end{align}
where $n_d=\langle{n_{\uparrow}}\rangle+\langle{n_{\downarrow}}\rangle$ is the local charge,
$m_d$ is the local moment, $\lambda_x$ and $\lambda_y$ are the Pauli matrices in the space of
two impurities, $\Sigma_2$ and $\Sigma_3$ are self-energy from hybridization between impurity
states and bulk electrons, the exact form of $\Sigma_2$ and $\Sigma_3$ are given by
\begin{align}
&\Sigma_2(i\omega_n)=\frac{12D\Gamma_0}{\pi{R}{k_T^3}}\int_{0}^{k_T}{\mathrm{d}}k
\frac{k(i\omega_n+\mu){\sin(kR)}}{(i\omega_n+\mu)^2-E_{\bm{k}}^2},\label{eq13}\\
&\Sigma_3(i\omega_n)=\frac{12D\Gamma_0}{\pi{R^2}{k_T^3}}\int_{0}^{k_T}{\mathrm{d}}k\frac{k[{\sin(kR)-kR\cos(kR)}]}{(i\omega_n+\mu)^2-E_{\bm{k}}^2},\label{eq14}\\
&E_{\bm{k}}^2=k^2+(1-mBk^2)^2. \label{eq15}
\end{align}
$\Sigma_2$ represents the normal propagation of electrons in the bulk of TIs, while $\Sigma_3$
represents the spin-flipping propagation of electrons in the bulk of TIs due to spin-orbit
coupling.
%In Eq. (\ref{eq12}), $\mathcal{U}$ is a SU(2) rotation matrix for each impurity site, which makes the direction of local moment $m_d$ well defined.
In Eq. (\ref{eq12}), we introduce a SU(2) matrix $\mathcal{U}$ to describe the spin degrees of freedom of the impurity with non-vanishing local moment.
$\mathcal{U}$ can be
parameterized in the following form,
\begin{equation}
{\cal U}=\left(\begin{array}{cc}b_1 & b_2^{\dagger} \\
-b_2 & b_1^{\dagger} \\
\end{array}\right),\label{eq16}
\end{equation}
with SU(2) constraint on the bosonic creation and annihilation operators $b^{\dagger}_{1,2}$, $b_{1,2}$:
\begin{equation}
b_1^{\dagger}b_1+b_2^{\dagger}b_2=1.\label{eq17}
\end{equation}
Local moment $m_d$ is defined as $m_d=\langle{f_{\uparrow}^{\dagger}f_{\uparrow}}\rangle-\langle{f_{\downarrow}^{\dagger}f_{\downarrow}}\rangle$ and $f=\mathcal{U}^{\dagger}d$.
Under the parametrization (\ref{eq16}) and up to the self-consistent Hartree-Fork approximation, we find that the local spin can be expressed as
\begin{align}
&S^{z}=d_{\uparrow}^\dagger{d_{\uparrow}}-d_{\downarrow}^\dagger{d_{\downarrow}}=m_d\left(b_1^\dagger{b_1}-b_2^\dagger{b_2}\right),\label{eq18}\\
&S^{x}=d_{\uparrow}^\dagger{d_{\downarrow}}+d_{\downarrow}^\dagger{d_{\uparrow}}=m_d\left(b_1^\dagger{b_2}+b_2^\dagger{b_1}\right),\label{eq19}\\
&S^{y}=-i\left(d_{\uparrow}^\dagger{d_{\downarrow}}-d_{\downarrow}^\dagger{d_{\uparrow}}\right)=-i{}m_d\left(b_1^\dagger{b_2}-b_2^\dagger{b_1}\right).\label{eq20}
\end{align}
One can find that Eq. (\ref{eq18})-(\ref{eq20}) tend to the Schwinger-Wigner representation\cite{Tsvelikbook} of spin when $m_d\rightarrow1$. In the loop approximation, we find that the spin-spin interaction between the two magnetic
impurities can be written as
\begin{equation}\label{eq21}
H_{RKKY}={\cal K}\bm{S}_{1}{\cdot}\bm{S}_{2}+{\cal Q}\left(S_{1}^{\parallel}S_{2}^{\parallel}-\bm{S}_{1}^{\perp}{\cdot}\bm{S}_{2}^{\perp}\right)+{\cal R}\left(\bm{S}_{1}{\times}\bm{S}_{2}\right)_{\parallel},
\end{equation}
where $\parallel$ ($\perp$) means the component(s) parallel (perpendicular) to the
y-axis. The range functions are given by
\begin{align}
&{\cal K}(R,\mu)=2J^2\frac{1}{\beta}\sum_{i\omega_n}\left[{\Sigma_2^2(i\omega_n)}\right],\label{eq22}\\
&{\cal Q}(R,\mu)=2J^2\frac{1}{\beta}\sum_{i\omega_n}\left[{\Sigma_3^2(i\omega_n)}\right],\label{eq23}\\
&{\cal R}(R,\mu)=2J^2\frac{1}{\beta}\sum_{i\omega_n}\left[2{\Sigma_2(i\omega_n)\Sigma_3(i\omega_n)}\right],\label{eq24}
\end{align}
and $J=U/(\epsilon_d-\mu+n_dU)^2$. The prefactor 2 indicates that there are two orbits in
the bulk of TIs. One can find that the spin-spin interaction between magnetic impurities
in the bulk of TI is similar to the RKKY interaction between magnetic impurities on TI
surface\cite{PhysRevLett.106.097201}. There are three different terms, an isotropic
Heisenberg-like term, an anisotropic term and a DM-like term. The difference is that the
anisotropic term is not an Ising-like term but a XXZ-model-like term.

Now we analyze the range functions in more details. It is difficult to get analytical
expressions due to the complex band structure of TIs, here we present the numerical results
for three typical cases: (1) the chemical potential is in the gap of TIs, $-1<\mu<1$, (2)
the chemical potential is in the conduction band, $\mu>1$, and (3) the chemical potential
is in the valance band, $\mu<-1$.

\begin{figure}[tb]
  \centering
  \includegraphics[width=\columnwidth]{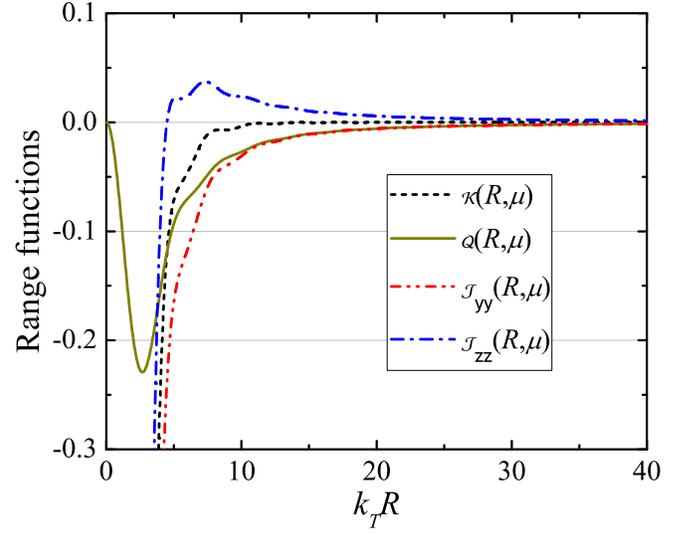}
  \caption{Range functions of RKKY interaction between two magnetic impurities
  as a function of distance $k_TR$ when chemical potential $\mu=0$.
  $\mathcal{J}_{yy}=\mathcal{K}(R,\mu)+\mathcal{Q}(R,\mu)$ is the longitudinal
  spin-spin interaction, $\mathcal{J}_{zz}=\mathcal{K}(R,\mu)-\mathcal{Q}(R,\mu)$
  is the transverse spin-spin interaction. Local moment $m_d=1$ and zero-temperature
  limit were taken in the calculation.}\label{fig4}
\end{figure}
\begin{figure}[tb]
  \centering
  \includegraphics[width=\columnwidth]{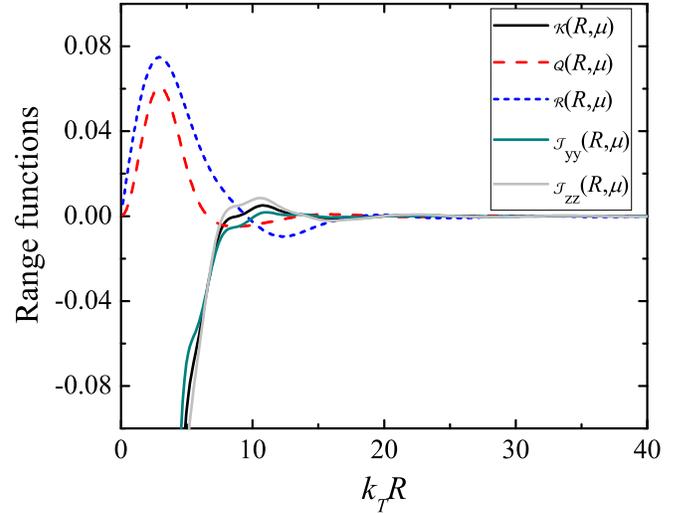}
  \caption{Range functions of RKKY interaction between two magnetic impurities
  as a function of distance $k_TR$ when chemical potential $\mu=2$.
  $\mathcal{J}_{yy}=\mathcal{K}(R,\mu)+\mathcal{Q}(R,\mu)$,
  $\mathcal{J}_{zz}=\mathcal{K}(R,\mu)-\mathcal{Q}(R,\mu)$ and
  $\mathcal{J}_{xz}=\mathcal{R}(R,\mu)$. Local moment $m_d=1$ and zero-temperature
  limit were taken in the calculation.}\label{fig5}
\end{figure}
\begin{figure}[tb]
  \centering
  \includegraphics[width=\columnwidth]{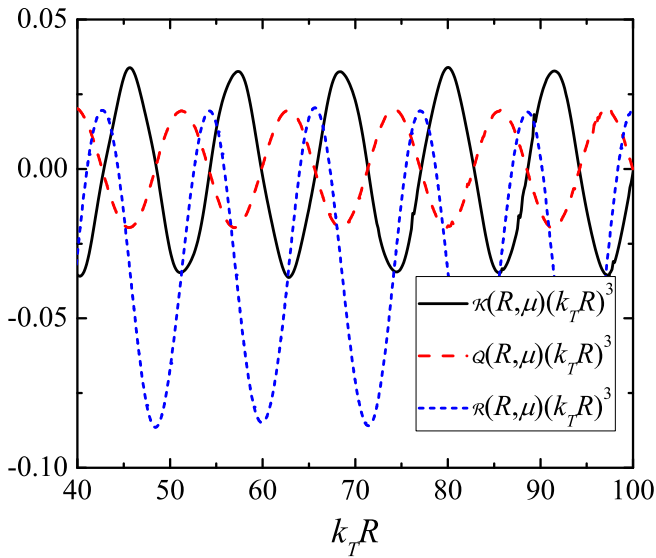}
  \caption{Product $\mathcal{K}(R,\mu){(k_TR)}^3$, $\mathcal{Q}(R,\mu){(k_TR)}^3$
  and $\mathcal{R}(R,\mu){(k_TR)}^3$ as functions of dimensionless distance $k_TR$.
  $\mu=2$, $m_d=1$ and $T=0$ were chosen in numerical calculation.}\label{fig6}
\end{figure}

For the first case, we find that the DM-like term $\mathcal{R}(R,\mu)$ equals to zero
spontaneously, and the other two terms decay exponentially with respect to $k_TR$ due to the BR interaction
mediated by massive Dirac electrons in the bulk of TIs. In the short distance limit
$R\rightarrow0$, the isotropic spin-spin interaction dominates, $\mathcal{K}(0,\mu)<0$,
and the anisotropic spin-spin interaction tends to vanish, $\mathcal{Q}(0,\mu)=0$. In the
long distance limit, we find that the anisotropic spin-spin interaction $\mathcal{Q}(R,\mu)$
dominates and the isotropic spin-spin interaction $\mathcal{K}(R,\mu)$ decays more rapidly
(one can see in Fig. (\ref{fig4}) that $\mathcal{J}_{yy}(R,\mu)=-\mathcal{J}_{zz}(R,\mu)$
in the long distance limit,{\it i.e.} $k_TR>10$). All of these phenomena are related to
spin-orbit coupling in TIs. When spin-orbit coupling tends to zero, the self-energy
$\Sigma_3\rightarrow0$ and $\mathcal{Q}(R,\mu)\rightarrow0$. A typical example is the short
distance limit. One can check in Eq. (\ref{eq14}) that $\Sigma_3\rightarrow0$ while
$R\rightarrow0$. The long distance limit is more interesting because it reflects the special
band structure of TIs. As shown in Fig. (\ref{fig1}), the minimal gap is not located at
$\bm{k}=0$, so that the decay length of spin-orbit coupled states are larger than that of
uncoupled states, and $\mathcal{K}(R,\mu)$ decays much more rapidly than $\mathcal{Q}(R,\mu)$.
Therefore, when the chemical potential is tuned into the gap of TIs, the diluted magnetic
impurities prefer paramagnetic phase not only because the effective interactions decay
exponentially but also because of the anisotropic spin-spin interaction dominates.

For the second and third cases, we find that $\mathcal{K}(R,\mu)=\mathcal{K}(R,-\mu)$,
$\mathcal{Q}(R,\mu)=\mathcal{Q}(R,-\mu)$ and $\mathcal{R}(R,\mu)=-\mathcal{R}(R,-\mu)$.
Shown in Fig. (\ref{fig5}) are the range functions for $\mu=2.0$. In the short distance limit
$R\rightarrow0$, we also find that
$\mathcal{Q}(R,\mu)=\mathcal{R}(R,\mu)=0$ and $\mathcal{K}(R,\mu)<0$. In short separation
range $0<k_TR<20$, the oscillation and decay are very complex due to the complex band
structure of TIs. In the large separation limit $R\rightarrow\infty$, the oscillation and
decay are determined by bulk states near the Fermi surface and as shown in
Fig. (\ref{fig6}), all of the three range functions decay with power law
$\sim\cos{(2k_FR)/R^3}$, which is the same as in the conventional 3D electron gas. There
are two interesting things in Fig (\ref{fig6}). Firstly, similar to the surface doping
\cite{PhysRevLett.106.097201}, one can see in Fig. (\ref{fig6}) that the red (long-dashed)
line and black (solid) line are almost of opposite sign, and the maxima and minima of blue
(short-dashed) line always appear at the zero points of red and black lines ({\it i.e.} $k_TR=60$),
where the Heisenberg-like term and XXZ-model-like term vanish and DM term dominates.
Secondly, the oscillation center of $\mathcal{R}(R,\mu)(k_TR)^3$ does not locate at zero,
which indicates that there is a remainder spin-rotation behind the oscillation.

In addition, we want to clarify that the range functions presented above tend to the
conventional RKKY interaction in the 3D electron gas in the following limits: spin-orbit
coupling $\rightarrow0$ and $\mu\gg1$. One can check that $\Sigma_3\rightarrow0$ when
spin-orbit coupling $\rightarrow0$, so that $\mathcal{Q}(R,\mu)=\mathcal{R}(R,\mu)=0$.
While $\mu\gg1$, the dispersion relation (\ref{eq15}) can be approximated by
$E_{\bm{k}}=\bm{k}^2/2M$ with $2M=1/mB$. We carry out the integration over momentum and
frequency in Eq. (\ref{eq13}) and Eq. (\ref{eq22}) in zero-temperature limit and obtain
\begin{equation}
\mathcal{K}(R,\mu)=\frac{J^2k_F^4}{2\pi{mB}}\left(\frac{12D\Gamma_0}{k_T^3}\right)^2\left(\frac{x\cos{x}-\sin{x}}{x^4}\right),\label{eq25}
\end{equation}
where $x=2k_FR$ and $k_F=\sqrt{\mu/mB}$. Eq. (\ref{eq25}) is exactly the conventional
RKKY interaction between magnetic impurities in 3D electron gas.

\section{Quantum Monte Carlo Simulations}\label{section3}

The mean-field results presented above shows that there are three different components in the
spin-spin interaction between magnetic impurities, and in long distance limit
($R\rightarrow\infty$), the interaction decays exponentially or with power law
$\sim1/R^3$ depending on the values of $\mu$. However, when quantum fluctuation is significantly large,
{\it i.e.} for intermediate values of $U$ or if the displacement between two impurities is very small ($R\rightarrow0$),
the mean-field results are suspectable. We need an unbiased method to study the spin-spin
interaction between magnetic impurities. In this section we report the numerical results of
spin-spin correlation functions obtained from the HFQMC simulations.

The Hirsch-Fye algorithm naturally returns the imaginary-time Green's functions
${\cal G}^{ss'}_{j,j'}(\tau,R)$ where $j,j'=1,2$ indicate two magnetic atoms and $s,
s'={\uparrow}, {\downarrow}$ are spin-indices. All of the information about the host material
is included in the input Green's functions ($U=0$) which can be obtained analytically, so
we can in principle deal with infinite host material in our QMC simulations.

From the imaginary-time Green's functions, we can carry out the trace over fermion variables
based on the Wick's theorem to calculate various correlation functions. For example, the
local moment squared on the impurity site is given by
\begin{equation}
m_d^2=\langle(n_{d\uparrow}-n_{d\downarrow})^2\rangle_{\{s\}},\label{eq26}
\end{equation}
where $\langle{...}\rangle_{\{s\}}$ denotes taking the average over discrete auxiliary field
introduced in the quantum Monte Carlo simulations.
The closer this value is to one, the more fully developed is the moment.

The displacement between the two magnetic impurities is along the y-direction, so the three
different types of the spin-spin correlations are defined as
\begin{align}
  &\langle{S_1^z}{S_2^z}\rangle_{\{s\}}=\langle{S_1^x}{S_2^x}\rangle_{\{s\}}  \nonumber\\
  &=\langle({\cal G}_{11}^{\uparrow\uparrow}-{\cal G}_{11}^{\downarrow\downarrow})\times({\cal G}_{22}^{\uparrow\uparrow}-{\cal G}_{22}^{\downarrow\downarrow})-{\cal G}_{12}^{\uparrow\uparrow}\cdot{\cal G}_{21}^{\uparrow\uparrow}-{\cal G}_{12}^{\downarrow\downarrow}\cdot {\cal G}_{21}^{\downarrow\downarrow} \nonumber\\
  &+{\cal G}_{12}^{\uparrow\downarrow}\cdot{\cal G}_{21}^{\downarrow\uparrow}+{\cal G}_{12}^{\downarrow\uparrow}\cdot{\cal G}_{21}^{\uparrow\downarrow}\rangle_{\{s\}},\label{eq27}\\
  & \langle{S_1^yS_2^y}\rangle_{\{s\}} \nonumber\\
  &= \langle-{\cal G}_{12}^{\uparrow\uparrow}\cdot{\cal G}_{21}^{\downarrow\downarrow}
  -{\cal G}_{12}^{\downarrow\downarrow}\cdot{\cal G}_{21}^{\uparrow\uparrow}
  +{\cal G}_{12}^{\uparrow\downarrow}\cdot{\cal G}_{21}^{\uparrow\downarrow}
  +{\cal G}_{12}^{\downarrow\uparrow}\cdot{\cal G}_{21}^{\downarrow\uparrow}\nonumber\\
  &-({\cal G}_{11}^{\uparrow\downarrow}-{\cal G}_{11}^{\downarrow\uparrow})\cdot ({\cal G}_{22}^{\uparrow\downarrow}-{\cal G}_{22}^{\downarrow\uparrow})\rangle_{\{s\}},\label{eq28}\\
  &\langle{S_1^xS_2^z}\rangle_{\{s\}}
  = \langle({\cal G}_{11}^{\uparrow\downarrow}+{\cal G}_{11}^{\downarrow\uparrow})\times({\cal G}_{22}^{\uparrow\uparrow}-{\cal G}_{22}^{\downarrow\downarrow})\nonumber\\
  & -{\cal G}_{12}^{\uparrow\uparrow}\cdot {\cal G}_{21}^{\uparrow\downarrow}
  +{\cal G}_{12}^{\downarrow\downarrow}\cdot {\cal G}_{21}^{\downarrow\uparrow}
  -{\cal G}_{12}^{\downarrow\uparrow}\cdot {\cal G}_{21}^{\uparrow\uparrow}
  +{\cal G}_{12}^{\uparrow\downarrow}\cdot {\cal G}_{21}^{\downarrow\downarrow}\rangle_{\{s\}}.\label{eq29}
\end{align}
In all the quantum Monte Carlo results we show below, we choose the on-site Coulomb repulsion
$U=3.0$ and the temperature $k_BT=1/16$ since the results for larger values of Coulomb
repulsion ($U=5.0$) and lower temperature ($T=1/32$) remain qualitatively unchanged.

\subsection{Local moment}\label{section3.1}

\begin{figure}[tb]
  \centering
  \includegraphics[width=\columnwidth]{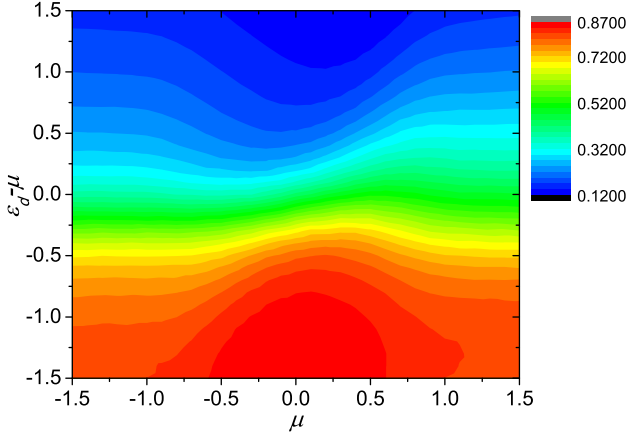}
  \caption{The HFQMC results of the local moment squared $\langle{m_d^2}\rangle$ for various values of
  chemical potential $\mu$ and impurity energy level $\epsilon_d$ when $k_TR=0$. We choose $U=3.0$ and
  the temperature is $k_BT=1/16$.}\label{fig7}
\end{figure}
\begin{figure}[tb]
  \centering
  \includegraphics[width=\columnwidth]{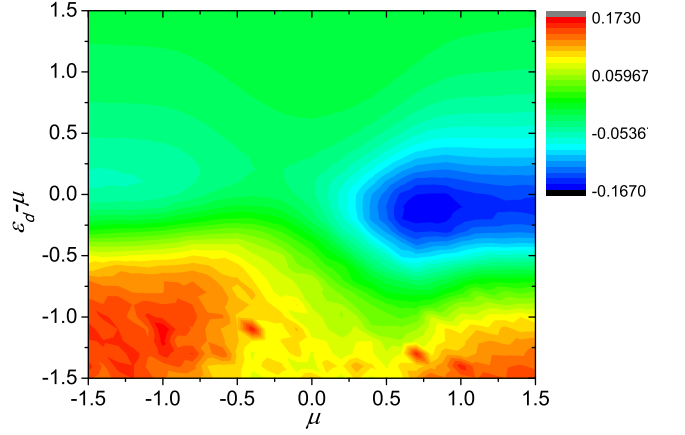}
  \caption{The HFQMC results of the spin-spin correlation for $k_TR=0$ for various values of chemical
  potential $\mu$ and impurity energy level $\epsilon_d$. We choose $U=3.0$ and the temperature is
  $T=1/16$.}\label{fig8}
\end{figure}

Following the guide of mean-field analysis, we firstly consider the local moment formation. Fig.
(\ref{fig7}) shows the local moment squared defined in Eq. (\ref{eq26}). The two magnetic
atoms are equally coupled to the TIs, so the local moment formed on the two impurity sites
shall be the same. We can see that when the impurity energy level is below the chemical
potential ($\epsilon_d<\mu$) a well-developed local moment is formed on the impurity site,
while for the case when $\epsilon_d>\mu$ the local moment has smaller values. Basing on the
fact that the impurity charge can either be zero or one, we note that
\begin{equation}\label{eq30}
m_d^2=n_d-2n_{d\uparrow}n_{d\downarrow},
\end{equation}
where ${n_d}={n_{d\uparrow}+n_{d\downarrow}}$.
When $\epsilon_d>\mu$, the impurity energy level is above the chemical potential, so the
charge number as well as the double occupancy drop drastically, and the decrease in the local
moment when $\epsilon_d>\mu$ is mainly caused by the decrease of charge number on the impurity
sites. However, when $\epsilon_d<\mu$ the impurity sites would prefer single occupancy when
$\epsilon_d+U>\mu$, so there would be a well-developed local moment formed on the impurities
in this parameter range. When $\epsilon_d<\mu$, we also find that the local moment is
preserved better when the chemical potential is lied in the gap of TIs since there are
no electrons to screen the local moment on the Fermi surface. In addition,
the  QMC results are qualitatively consistent with self-consistent Hartree-Fork approximation,
which demonstrates that the mean-field results for single impurity in the bulk of TIs are valid (see
Fig. (\ref{fig7}) and Fig. (\ref{fig3})).

\subsection{Spin-Spin correlation functions}\label{section3.2}

\begin{figure*}[tb]
  \centering
  \includegraphics[width=\textwidth]{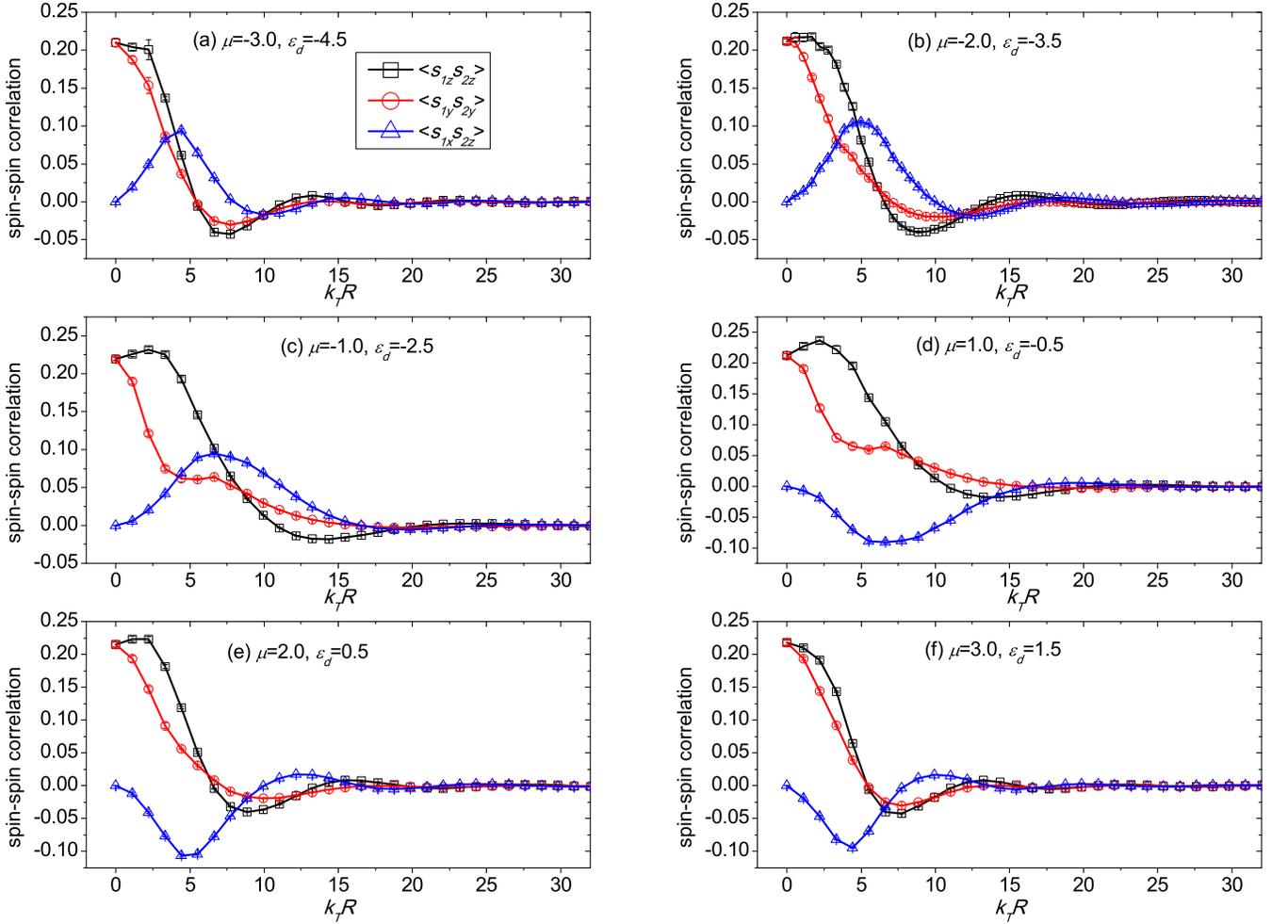}
  \caption{The results of spin-spin correlation with respect to the displacement $k_TR$ for the cases
  that $\epsilon_d<\mu$. (a) $\mu=-3.0$, $\epsilon_d=-4.5$, (b) $\mu=-2.0$, $\epsilon_d=-3.5$, (c)
  $\mu=-1.0$, $\epsilon_d=-2.5$, (d) $\mu=1.0$, $\epsilon_d=-0.5$, (e) $\mu=2.0$, $\epsilon_d=0.5$, (f)
  $\mu=3.0$, $\epsilon_d=1.5$.}\label{fig9}
\end{figure*}

In QMC simulations, the two-impurity spin-spin correlation
functions defined in Eq. (\ref{eq27})-(\ref{eq29}) are proper quantities to describe the
interaction between magnetic impurities. For the weak coupling case where Eq. (\ref{eq21})
can be used, one can check that the spin-spin correlation functions can be written as
\begin{align}
&\langle{S_1^{y}S_2^{y}}\rangle=
\frac{e^{-\beta\mathcal{J}_{\parallel}}-e^{\beta\mathcal{J}_{\parallel}}\cosh(2\beta\mathcal{J}_{\perp})}
{e^{-\beta\mathcal{J}_{\parallel}}+e^{\beta\mathcal{J}_{\parallel}}\cosh(2\beta\mathcal{J}_{\perp})}m_d^2,\\
&\langle{S_1^{x}S_2^{x}}\rangle=-\frac{\mathcal{K}-\mathcal{Q}}{\mathcal{J}_{\perp}}
\frac{e^{\beta\mathcal{J}_{\parallel}}\sinh(2\beta\mathcal{J}_{\perp})}
{e^{-\beta\mathcal{J}_{\parallel}}+e^{\beta\mathcal{J}_{\parallel}}\cosh(2\beta\mathcal{J}_{\perp})}m_d^2,\\
&\langle{S_1^{x}S_2^{z}}\rangle=-\frac{\mathcal{R}}{\mathcal{J}_{\perp}}
\frac{e^{\beta\mathcal{J}_{\parallel}}\sinh(2\beta\mathcal{J}_{\perp})}
{e^{-\beta\mathcal{J}_{\parallel}}+e^{\beta\mathcal{J}_{\parallel}}\cosh(2\beta\mathcal{J}_{\perp})}m_d^2,
\end{align}
where $\mathcal{J}_{\parallel}=\mathcal{K}+\mathcal{Q}$,
$\mathcal{J}_{\perp}=\sqrt{(\mathcal{K}-\mathcal{Q})^2+\mathcal{R}^2}$.
In the low temperature region where $\beta\mathcal{K}, \beta\mathcal{Q}$ and $\beta\mathcal{R}\ll1$, the spin-spin
correlation functions are proportional to the spin-spin interactions,
$\langle{S_1^{y}S_2^{y}}\rangle\sim-\beta(\mathcal{K}+\mathcal{Q})m_d^2$,
$\langle{S_1^{x}S_2^{x}}\rangle\sim-\beta(\mathcal{K}-\mathcal{Q})m_d^2$,
$\langle{S_1^{x}S_2^{z}}\rangle\sim\beta\mathcal{R}m_d^2$.

Before a more detailed discussion of general spin-spin correlations, we consider the special
case $k_TR=0$ firstly. Mean-field analysis demonstrate that the anisotropic and the DM-like
spin-spin interactions vanishes in this special case. In QMC simulation, we find similar
results, the longitudinal and transverse components of the spin-spin correlation defined
in Eq. (\ref{eq27}) and Eq. (\ref{eq28}) have the same values, and the DM-like component
vanishes. Fig. (\ref{fig8}) shows the numerical results of isotropic spin-spin correlation
function. Two issues need to be addressed about Fig. (\ref{fig8}). Firstly, when the chemical
potential is tuned into the gap ({\it i.e.} $-0.5<\mu<0.5$), the correlation between
magnetic impurities is suppressed. This is because the correlation is mediated by itinerant
electrons. Secondly, the spin-spin correlation can either be FM or AFM, while the mean-field
analysis prefers FM correlation for any given parameters (as shown in Fig. (\ref{fig4}) and
Fig. (\ref{fig5}), $\mathcal{K}(0,\mu)<0$).

Now we present the spin-spin correlation with respect to the displacement $k_TR$ for some
typical values of $\epsilon_d$ and $\mu$ (Fig. (\ref{fig9})). In all the cases we test,
we fix $\epsilon_d-\mu=-1.5$ which most prefer
local moment formation when the on-site Coulomb repulsion $U=3.0$. In Fig. (\ref{fig9}a)-(\ref{fig9}c),
we present the results of spin-spin correlation for the cases that the chemical potential is
tuned into the valence band. Firstly, similar to the mean-field results, the DM-like
correlation $\langle{S_1^xS_2^z}\rangle$ and the difference between the longitudinal and
transverse correlations tend to zero smoothly in short separation limit. Secondly, the
difference between the longitudinal and transverse correlations becomes more apparent as
the chemical potential approaches zero. Shown in Fig. (\ref{fig9}d)-(\ref{fig9}f) are the
results of spin-spin correlation when $\mu$ is switched into the conduction band. The FM
behavior of the longitudinal and transverse correlations in the short displacement range
remain unchanged. However, we can see that the oscillation of $\langle{S_1^xS_2^z}\rangle$
are opposite, which is qualitatively consistent with the mean-field results
$\mathcal{K}(R,\mu)=\mathcal{K}(R,-\mu)$, $\mathcal{Q}(R,\mu)=\mathcal{Q}(R,-\mu)$ and
$\mathcal{R}(R,\mu)=-\mathcal{R}(R,-\mu)$.

\begin{figure}[tb]
  \centering
  \includegraphics[width=\columnwidth]{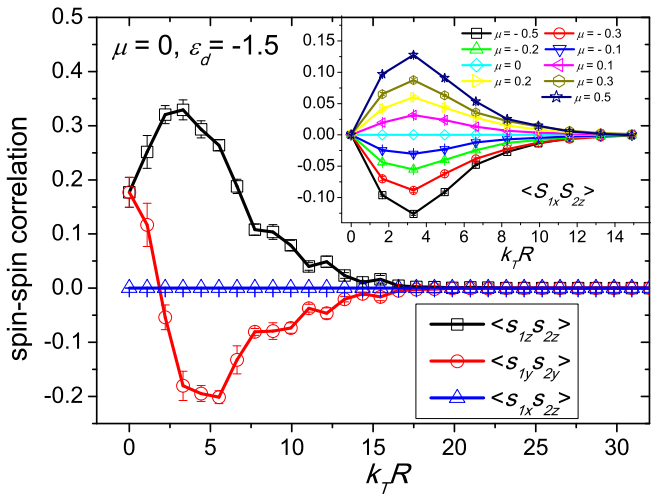}
  \caption{The results of spin-spin correlation with respect to the displacement $k_TR$ for $\mu=0$ and
  $\epsilon_d=-1.5$. The inset shows the results of DM-like spin-spin correlation for various values of
  $\mu$ while $\epsilon_d-\mu=-1.5=-U/2$.}\label{fig10}
\end{figure}

In Fig. (\ref{fig10}) we show the results of spin-spin correlation when the chemical potential
is tuned into the gap of the TIs ($\mu=0$) for $\epsilon_d=-1.5$. The first important feature
in this figure is that the transverse DM-like correlation is vanished.
Theoretically, one can demonstrate that $\langle{S_1^xS_2^z}\rangle$ is exactly zero according
to the particle-hole symmetry ($\Psi\rightarrow\tau_y\sigma_y\Psi^{\dagger}$,
$d\rightarrow\sigma_y{d}^\dagger$) and inversion symmetry ($\Psi\rightarrow\tau_z\Psi$,
$d_{1}\leftrightarrow{d}_2$). Actually, under the combined symmetric transformation, one
can find that $\langle{S_1^xS_2^z}\rangle=\langle{S_2^xS_1^z}\rangle$, however,
$\langle{S_1^xS_2^z}\rangle=-\langle{S_1^zS_2^x}\rangle$ according to rotation symmetry, so
$\langle{S_1^xS_2^z}\rangle=0$.
According to our QMC calculations, the values of the DM-like correlation is exactly zero when
$\mu=0$ and $\epsilon_d=-1.5=-U/2$, which means that our QMC simulations preserve the symmetries exactly.
We also find that as the chemical potential approaches zero, the amplitude
of the DM-like correlation function decreases for both positive and negative values of $\mu$,
as shown in the inset of Fig. (\ref{fig10}).
The second important feature in Fig. (\ref{fig10}) is that the distinction between
$\langle{S_1^zS_2^z}\rangle$ and $\langle{S_1^yS_2^y}\rangle$ is significant when $\mu=0$.
One can find that such a distinction is more remarkable than that in the mean-field results (see Fig.
(\ref{fig4})).

\section{Discussion and conclusion}\label{section4}

In this work, we apply mean-field theory and HFQMC method to study the spin-spin
interaction in the bulk of TIs. We find that the spin-spin interaction has three different components: the longitudinal,
the transverse and the transverse DM-like interaction induced by
the spin-orbit coupling. From the mean-field calculation, we find that for a heavy doped
system, all three kinds of spin-spin interaction oscillate with the same period
$2\pi/\sqrt{|\mu|/mB}$, and decay with $\sim1/R^3$ in the limit $R\rightarrow\infty$.
Both the quantum Monte Carlo simulations and mean-field calculations demonstrate that the longitudinal
interaction oscillates like the transverse one, and both of them are
always ferromagnetic for $\epsilon_d<\mu$ in the short range limit, ${k_T}R\rightarrow0$.
When the Fermi energy is located in the bulk gap of TIs, the spin-spin interaction
decays exponentially due to the Bloembergen-Rowland interaction. The mean-field calculation
and the HFQMC simulation demonstrate that (For sufficiently large distance, {\it i.e.} $k_TR>5$) the
longitudinal interaction is antiferromagnetic and transverse ones are ferromagnetic.

For the real material Fe$_x$Bi$_2$Te$_3$, if we ignore the influence of anisotropic crystal
structure and choose the following parameters{\cite{PhysRevB.82.045122}}: lattice spacing
$a=4.38${\AA}, $v_F=2.87$eV{\AA} and $m=0.3$eV, then $v_F{k_T}/m\thickapprox5.53$ and
${k_T}a\thickapprox2.53$. $k_T{R}\sim10$ corresponds to $R{\sim}4a$ and concentration of
impurity $x{\sim}0.016$ which defines a typical length scale. When $k_T{R}<10$, the
difference between longitudinal correlation and transverse correlation and the importance
of DM-like correlation become significant (see Fig. (\ref{fig9}) and Fig. (\ref{fig5})).
However, more detailed analysis, such as the anisotropic lattice structure, the position
of magnetic impurities, are required to explain the topological phase transition between
ferromagnetic-cluster glassy behavior for $x\sim0.025$ and valence-bond glassy behavior
for $x\sim0.03$ to $0.1$.

\section{Acknowledgement}
This work is supported by NSAF (Grant No. U1230202), China Postdoctoral Science Foundation
(Grant No. 2013M540845), and CAEP.


%\appendix
%\section{Derivation of spin-spin interaction}

%-----------------------------------------------------------------
% Sec**: References
%-----------------------------------------------------------------
%\nocite{*}
\bibliographystyle{apsrev4-1}
\bibliography{ref}

\end{document}